\documentclass[wsdraft]{ws-procs961x669}  
 \begin{document}
\title{Transition Distribution Amplitudes : from JLab to EIC }
\author{L. Szymanowski$^*$}
\address{NCBJ, 02-093 Warsaw, Poland\\
$^*$E-mail: Lech.Szymanowski@ncbj.gov.pl }
\author{B. Pire}
\address{CPHT, CNRS, \'Ecole polytechnique, I.P. Paris, 91128-Palaiseau, France}
\author{K.~Semenov-Tian-Shansky}
\address{National Research Centre Kurchatov Institute, Petersburg Nuclear Physics Institute, RU-188300 Gatchina, Russia}
\begin{abstract}
Baryon-to-meson Transition Distribution Amplitudes extend both the concepts of generalized parton
distributions  and baryon distribution amplitudes  encoding
 valuable complementary information on the $3$-dimensional hadronic structure. The recent
 analysis of backward meson electroproduction at JLab supports the hope to perform femto-photography of hadrons from a brand new perspective.and in particular opens a new domain for EIC experiments.
\end{abstract}
\keywords{QCD, TDA, EIC}
\bodymatter
\section{Introduction}
Baryon-to-meson Transition Distribution Amplitudes~\cite{Pire:2016aqa}  (TDAs) are matrix elements of a three quark operator between a baryon and a meson states.  
Since the corresponding operator carries the quantum numbers of a baryon,
baryon-to-meson TDAs allow the exploration of the  baryonic content of a nucleon.
This is complementary to the information one can obtain from  generalized parton distributions (GPDs), with the operator carrying the quantum numbers of mesons. 
Similarly to the case of GPDs, by Fourier transforming baryon-to-meson TDAs to the
impact parameter space,
one obtains additional insight on the baryon structure in the transverse plane.
The nonlocal three quark (antiquark) operator
on the light cone
($n^2=0$) occurring in the definition of  baryon-to-meson  TDAs reads (gauge links are omitted):
\begin{equation}
\hat{O}^{\alpha \beta \gamma}_{\rho \tau \chi}( \lambda_1 n,\, \lambda_2 n, \, \lambda_3 n)
 =
\varepsilon_{c_{1} c_{2} c_{3}}
\Psi^{c_1 \alpha}_\rho(\lambda_1 n)
\Psi^{c_2 \beta}_\tau(\lambda_2 n)
\Psi^{c_3 \gamma}_\chi (\lambda_3 n),
\label{operators}
\end{equation}
where
$\alpha$, $\beta$, $\gamma$
stand for the quark (antiquark) flavor indices,
$\rho$, $\tau$, $\chi$
denote the Dirac spinor indices and $c_i$ stand for the color indices. Baryon-to-meson  TDAs~
\cite{ Pire:2004ie,Pire:2005ax},
share common features both with baryon distribution amplitudes (DAs), introduced as baryon-to-vacuum matrix elements of the same operators,
and with generalized parton distributions (GPDs), since the matrix element in question depends on the longitudinal momentum transfer
$\Delta^+=(p_{\cal M}-p_1) \cdot n$
between a baryon 
$N(p_1)$ 
and a meson 
${\cal M}({p_{\cal M}})$ 
characterized by the skewness variable
$
\xi= -\frac{(p_{\cal M}-p_1) \cdot n}{(p_{\cal M}+p_1) \cdot n}
$
and by the transverse momentum transfer
$\vec \Delta_T$.
 \begin{figure}[h!]
\center
\includegraphics[width=0.3\columnwidth]{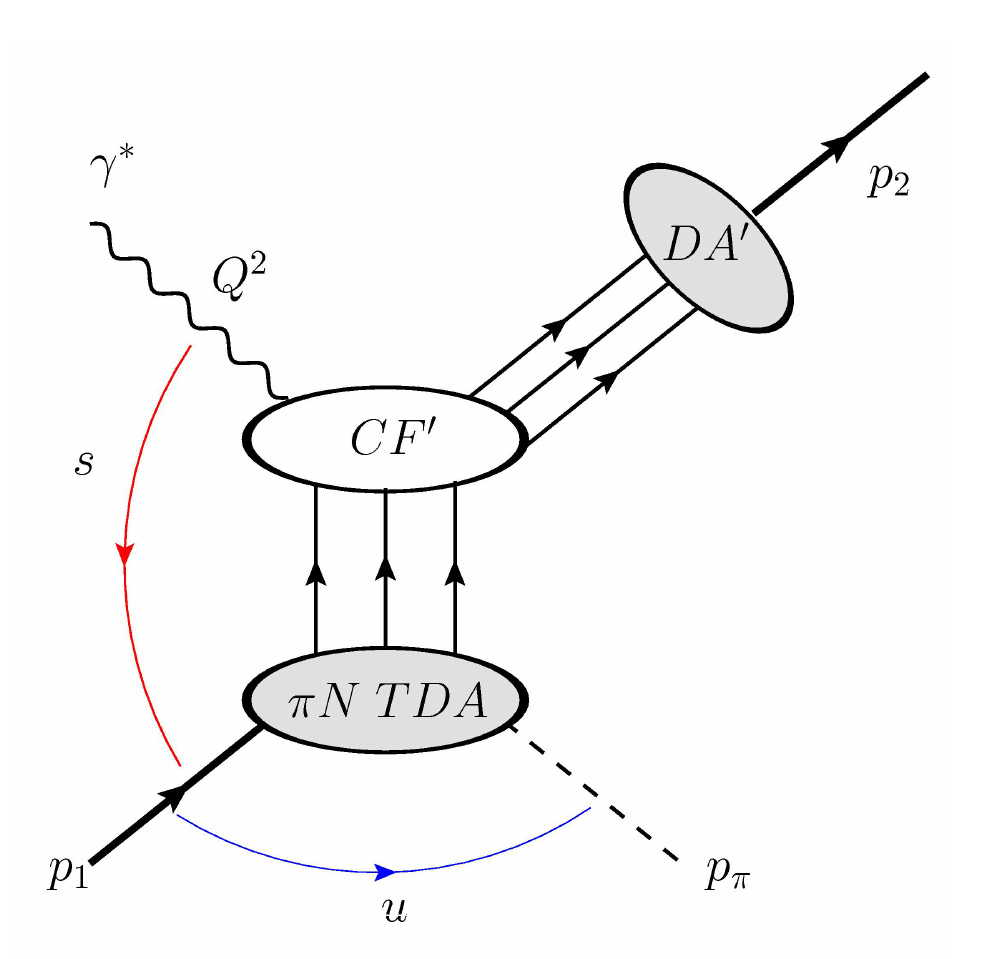}
\caption{ At large $Q^2=-q^2$ for the near-backward angles ({\it i.e.} for small $-u=-(p_1-p_\pi)^2$),
the $\gamma^*(q) N(p_1)\to \pi(p_{\pi}) N'(p_2)$ amplitude factorizes into a convolution of
a coefficient function (CF) with a baryon DA  and a baryon-to-meson TDA.  }
   \label{Fig_1}
\end{figure}
The collinear QCD factorization property, which ensures the validity of the 
GPD-based description of hard exclusive near-forward meson electroproduction reactions, 
may be extended to the complementary near-backward kinematical regime provided that the
$\bar{\Psi} \Psi$ operator is replaced by 
$\hat{O}^{\alpha \beta \gamma}_{\rho \tau \chi}( \lambda_1 n,\, \lambda_2 n, \, \lambda_3 n)$.
This results in the reaction mechanism sketched in Fig.~\ref{Fig_1} with a 
GPD replaced by baryon-to-meson TDA and baryon DA replacing meson DA.    

\section{The physical picture}

\begin{figure}[h!]
\center
\includegraphics[width=0.9\columnwidth]{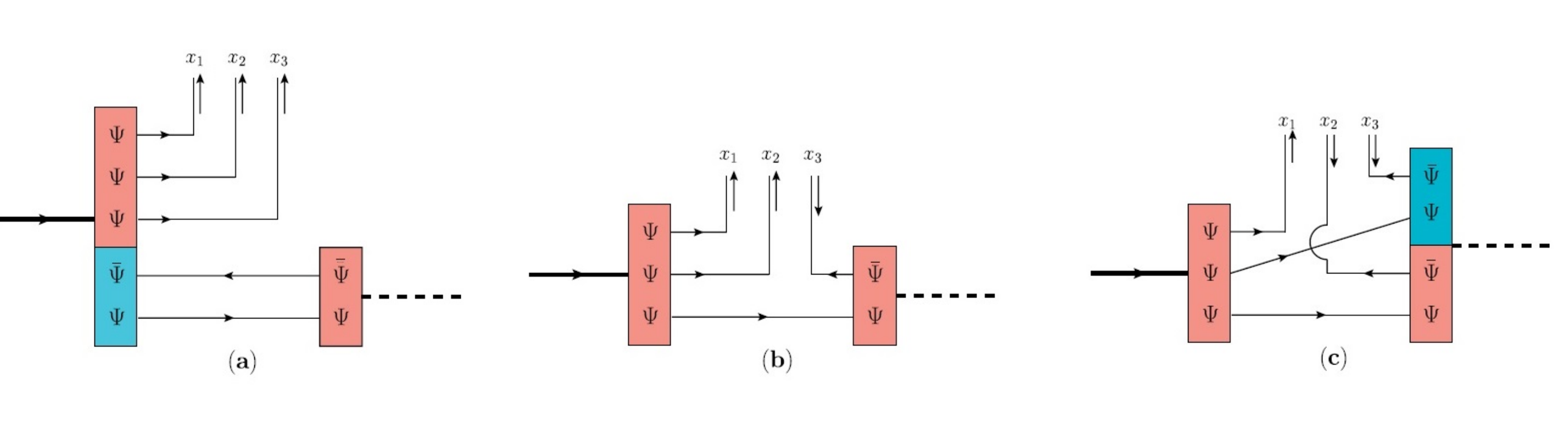}
\caption{Interpretation of baryon-to-meson TDAs at low normalization scale.
    Small vertical arrows show the flow of the momentum.
     {\bf  (a):} Contribution in the ERBL region (all $x_i$ are positive);
    {\bf  (b):} Contribution in the DGLAP I region (one of $x_i$  is negative).
    {\bf  (c):} Contribution in the DGLAP II region (two  $x_i$  are negative).}
   \label{Fig_X}
\end{figure}

The physical picture encoded in baryon-to-meson TDAs is conceptually close to that
contained in baryon GPDs and baryon DAs. Baryon-to-meson TDAs  characterize partonic correlations inside a baryon and give access to the momentum distribution
of the baryonic number inside a baryon. The same operator also defines the nucleon DA,
which can be seen as a limiting case of baryon-to-meson TDAs with the meson state replaced by the vacuum.
In the language of the Fock state decomposition, the leading twist baryon-to-meson TDAs are not restricted to the lowest Fock state
as the leading twist DAs. They rather probe the non-minimal Fock components with an additional
quark-antiquark pair:
\begin{eqnarray}
&&
| {\rm Nucleon} \rangle= |\Psi \Psi \Psi \rangle+ |\Psi \Psi \Psi; \,  \bar{\Psi} \Psi \rangle+....\;; ~~~~
| {\cal M} \rangle= |\bar{\Psi}\Psi \rangle+ |\bar{\Psi}\Psi; \, \bar{\Psi} \Psi \rangle+....\;
\end{eqnarray}
depending on the particular support region in question
(see Fig.~\ref{Fig_X}).

\section{JLab lessons}
\begin{figure}[h!]
\center
\includegraphics[width=0.3\columnwidth]{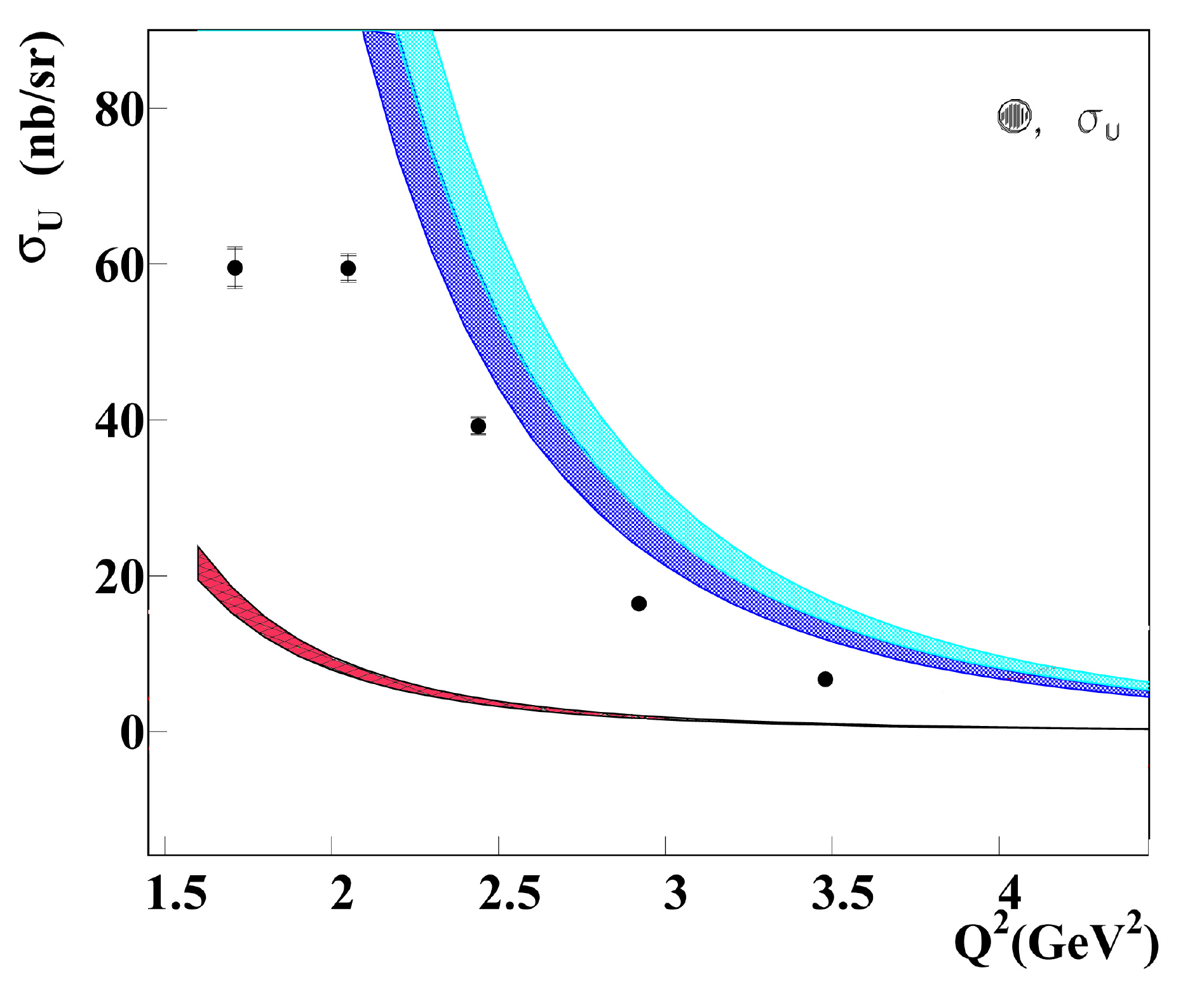} ~~~~~~~~~\includegraphics[width=0.35\columnwidth]{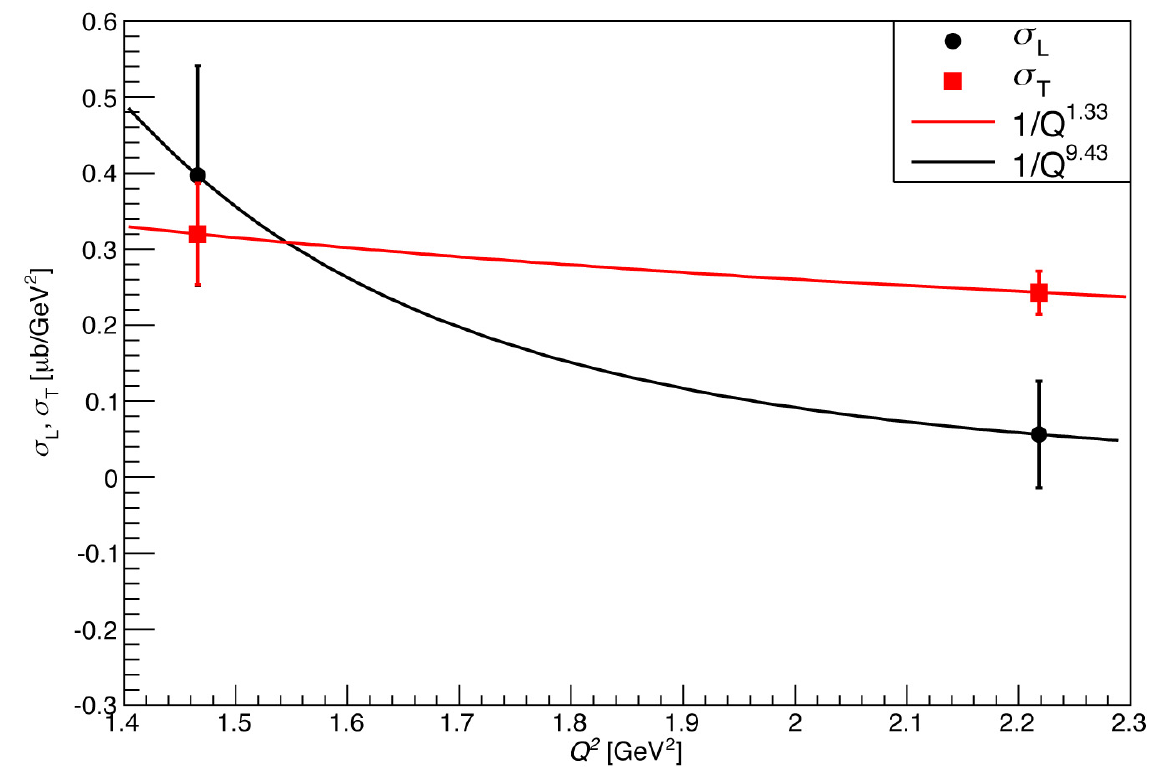}
\caption{Left panel: The $Q^2$ dependence of the $(\gamma^* p \to \pi^+ n)$ backward cross section $\sigma_U = \sigma_T+ \varepsilon \sigma_L$ measured by CLAS \cite{Park:2017irz} (circles) compared to the QCD prediction (colored bands) with three models of TDAs. Right panel: The $Q^2$ dependence of the longitudinal (squares)  and transverse (circles)  cross sections $\sigma_{L,T}(\gamma^* p \to \omega p)$ (in $\mu b$ GeV$^{-2}$) as measured at Hall C in JLab~\cite{Li-Huber}.} 
   \label{Fig_CLAS-Bill}
\end{figure}

The first experimental indications of the validity of the TDA concept have been recently 
presented~\cite{Park:2017irz, Li-Huber}. 
The left panel of Fig.~\ref{Fig_CLAS-Bill} shows the results of the $Q^2$-dependence of $\sigma_T + \epsilon~\sigma_L$ ($\varepsilon$ is the virtual photon linear polarization parameter) obtained by the CLAS collaboration at JLab for the  $e p \to e^\prime n \pi^+$ reaction in relatively large $Q^2$ ($>1.7$ GeV$^2$) and
small-$|u|$ domain ($\langle -u \rangle =0.5$ GeV$^2$) above the resonance region ($W^2=(q+p_1)^2> 4\,{\rm GeV}^2$), {\it i.e.} close to the ``near backward'' kinematical
regime  in which the TDA formalism is potentially applicable.
As anticipated, the cross sections has a strong $Q^2$-dependence.
The data are compared to the theoretical predictions of $\sigma_T$
from the cross channel nucleon pole exchange $\pi N$ TDA model suggested in Ref.~\cite{Pire:2011xv}.
The curves show the results of three  theoretical calculations
using different input phenomenological solutions for nucleon DAs
with their uncertainties represented by the bands. The crucial point is that the TDA formalism involves a dominance of the transverse amplitude.
Therefore, in order to be able to claim the validity of the TDA approach it
is necessary to separate $\sigma_T$ from $\sigma_L$  and check
that $\sigma_T \gg \sigma_L$.
This goal has been fulfilled by the Hall C experiment~\cite{Li-Huber} at JLab, which measured the reaction $e p \to e^\prime p^\prime \omega$ in a similar kinematic range. The data shown on  the right panel of  Fig.~\ref{Fig_CLAS-Bill} show indeed that $\sigma_T$ dominates over  $\sigma_L$ for sufficiently large values of $Q^2$, as anticipated by the collinear QCD factorization approach~\cite{Pire:2015kxa}.

\section{Perspectives for EIC}
TDAs are opening a new window on the study of the $3$-dimensional structure of nucleons
and recent experimental analysis of  backward meson electroproduction  hints that this concept may be applicable at moderate values of $Q^2$.
Clearly a more detailed experimental analysis  is required for the confirmation of  relevance of TDAs concept for analysis of hard processes. Testing the validity of the collinear factorized description of hard backward meson electroproduction reactions at the energies of Jlab@12 GeV  will help to elaborate a unified and consistent approach for hard exclusive reactions. Moreover, backward DVCS is also a very interesting channel to be explored; it is  a source of new information on the D-term form factor analytically continued to large $-t$. Let us stress also that TDAs are natural concepts to be used for the description of nuclear break-up reaction - such as deuteron electrodissociation - which may be interesting to visualize the partonic content of light nuclei. Let us also note that  the \={P}ANDA experiment at GSI-FAIR~\cite{Lutz:2009ff} will provide opportunities to access the cross conjugated counterparts of the reaction depicted on Fig. 1 and test the universality of baryon-to-meson 
TDAs~\cite{Lansberg:2012ha, Pire:2013jva}. 

Although detailed predictions have not yet been worked out for higher energies, one can anticipate that studies at the electron-ion collider (EIC) will allow this new domain physics to be further explored. Higher $Q^2$ should be accessible in a domain of moderate $\gamma^* N$ energies, i.e. rather small values of the usual $y$ variable and not too small values of $\xi$. The peculiar EIC kinematics, as compared to fixed target experiments, allows in principle a thorough analysis of the backward region pertinent to TDA studies. More phenomenological prospective studies are clearly needed.

\section*{Acknowledgments}
 L. S. is supported by the grant 2017/26/M/ST2/01074 of the National Science Center in Poland. He thanks the LABEX P2IO and the  GDR QCD for support. 

\end{document}